\documentclass[twocolumn,showpacs,aps,prl,superscriptaddress,floatfix,preprintnumbers]{revtex4}

\usepackage{graphicx}
\usepackage{dcolumn}
\usepackage{amsmath}
\usepackage{epsfig}

\RequirePackage{xspace}
\usepackage{relsize}
\def\babar{\mbox{\slshape B\kern-0.1em{\smaller A}\kern-0.1em
    B\kern-0.1em{\smaller A\kern-0.2em R}}}
\def\CP                {\ensuremath{C\!P}\xspace}
\def\en         {\ensuremath{e^-}\xspace}
\def\ep         {\ensuremath{e^+}\xspace}
\def\Dbar    {\kern 0.2em\overline{\kern -0.2em D}{}\xspace}
\def\Dz      {\ensuremath{D^0}\xspace}
\def\Dzb     {\ensuremath{\Dbar^0}\xspace}
\def\Dstarp  {\ensuremath{D^{*+}}\xspace}

\def\Y#1S{\ensuremath{\Upsilon{(#1S)}}\xspace}% no space before {...}!

\def\invfb   {\ensuremath{\mbox{\,fb}^{-1}}\xspace}
\newcommand{\mev}{\ensuremath{\mathrm{\,Me\kern -0.1em V}}\xspace}
\newcommand{\mevc}{\ensuremath{{\mathrm{\,Me\kern -0.1em V\!/}c}}\xspace}
\newcommand{\mevcc}{\ensuremath{{\mathrm{\,Me\kern -0.1em V\!/}c^2}}\xspace}
\newcommand{\gev}{\ensuremath{\mathrm{\,Ge\kern -0.1em V}}\xspace}
\newcommand{\gevc}{\ensuremath{{\mathrm{\,Ge\kern -0.1em V\!/}c}}\xspace}
\newcommand{\gevcc}{\ensuremath{{\mathrm{\,Ge\kern -0.1em V\!/}c^2}}\xspace}

\def\pep2{PEP-II}

\newcommand{\dzkpi}{\mbox{\ensuremath{\Dz \to K^{-} \pi^{+}}}}
\newcommand{\dzkpiws}{\mbox{\ensuremath{\Dz \to K^{+} \pi^{-}}}}
\newcommand{\dzbkpi}{\mbox{\ensuremath{\Dzb \to K^{+} \pi^{-}}}}
\newcommand{\dzkk}{\mbox{\ensuremath{\Dz \to K^{-} K^{+}}}}
\newcommand{\dzbkk}{\mbox{\ensuremath{\Dzb \to K^{+} K^{-}}}}
\newcommand{\dzpipi}{\mbox{\ensuremath{\Dz \to \pi^{-} \pi^{+}}}}
\newcommand{\dzbpipi}{\mbox{\ensuremath{\Dzb \to \pi^{+} \pi^{-}}}}
\newcommand{\spi}{\ensuremath{\pi_s}}
\newcommand{\dm}{\ensuremath{\Delta m}}

\newcommand{\BABARPubYear}     {07}
\newcommand{\BABARPubNumber}  {058}
\newcommand{\SLACPubNumber} {12756}

\begin{document}

\preprint{\babar-PUB-\BABARPubYear/\BABARPubNumber, SLAC-PUB-\SLACPubNumber} 

\title{\boldmath Search for \CP\ Violation in the Decays \dzkk\ and \dzpipi}

%% author list as of 05-Jul-2007 (565 authors)
%
\author{B.~Aubert}
\author{M.~Bona}
\author{D.~Boutigny}
\author{Y.~Karyotakis}
\author{J.~P.~Lees}
\author{V.~Poireau}
\author{X.~Prudent}
\author{V.~Tisserand}
\author{A.~Zghiche}
\affiliation{Laboratoire de Physique des Particules, IN2P3/CNRS et Universit\'e de Savoie, F-74941 Annecy-Le-Vieux, France }
\author{J.~Garra~Tico}
\author{E.~Grauges}
\affiliation{Universitat de Barcelona, Facultat de Fisica, Departament ECM, E-08028 Barcelona, Spain }
% ...
\collaboration{The \babar\ Collaboration}
\noaffiliation

\date{\today}

\begin{abstract}
We measure $C\!P$-violating asymmetries of neutral
charmed mesons in the modes $D^0 \to K^- K^+$
and $D^0 \to \pi^- \pi^+$ with the highest precision to date by using 
$D^0 \to K^- \pi^+$ decays to correct detector asymmetries. An analysis
of 385.8\,fb$^{-1}$ of data
collected with the BaBar detector yields values of
$a_{C\!P}^{K\!K} = (0.00 \pm 0.34\,\textrm{(stat.)} \pm 0.13\,\textrm{(syst.)})\%$
and
$a_{C\!P}^{\pi\pi} = (-0.24 \pm 0.52\,\textrm{(stat.)} \pm 0.22\,\textrm{(syst.)})\%$,
which agree with Standard Model predictions.

\end{abstract}

\pacs{14.40.Lb, 13.25.Ft, 11.30.Er}

% http://www.aip.org/pacs/pacs06/pacs0610.html

\maketitle

%%%%%%%%%%%%%%%%%%%%%%%%
%% BODY
%%%%%%%%%%%%%%%%%%%%%%%%

Charge-parity (\CP) violation in decays of charmed mesons at levels as large
as 1\% has not yet been experimentally ruled out~\cite{expresults},
and at this level would be evidence of unknown physical
phenomena~\cite{Grossman:2006jg,newmodels}.
The \CP-even decays \dzkk\ and \dzpipi~\cite{conjugates} are Cabibbo suppressed,
with the two neutral charmed mesons, \Dz\ and \Dzb, sharing the final states.
\CP-violating asymmetries in these modes 
are predicted to be $\mathcal{O}(0.001\%\textrm{--}0.01\%)$
in the Standard Model of particle physics~\cite{standardmodel},
yet have not been measured precisely due to limited sample sizes and
relatively large systematic effects~\cite{rpp}.

We search for \CP\ violation in decays of charmed mesons produced 
from charm-quark pairs in the reaction $e^+ e^- \to c\bar{c}$ by measuring the asymmetries
in the partial decay widths, $\Gamma$,
\begin{eqnarray}
& \displaystyle a_{\CP}^{K\!K}=\frac{\Gamma(\dzkk)-\Gamma(\dzbkk)}{\Gamma(\dzkk) + \Gamma(\dzbkk)} & \\
& \displaystyle a_{\CP}^{\pi\pi}=\frac{\Gamma(\dzpipi)-\Gamma(\dzbpipi)}{\Gamma(\dzpipi)+\Gamma(\dzbpipi)}\textrm{.} &
\end{eqnarray}
In this construction, $a_{\CP}^{hh}$, $h = K,\pi$, includes all \CP\ violating contributions,
direct and indirect~\cite{Grossman:2006jg}.  The presence of direct \CP\ violation in
one or both modes would be signaled by a non-vanishing difference between the modes,
$a_{\CP}^{K\!K} - a_{\CP}^{\pi\pi} \neq 0$.

Precise quantification of asymmetry in \Dz-flavor assignment, called tagging,
has long been considered the primary experimental challenge in
these measurements.  We develop a new
technique for measuring and correcting this asymmetry using only the recorded data.
However, forward-backward (FB) asymmetry in
$c\bar{c}$ production may be more significant at the center-of-mass energy
of $e^+ e^-$ collisions in \babar, $\sqrt{s} \approx 10.6\gev$.
This production asymmetry will create a difference in the numbers of
reconstructed \Dz\ and \Dzb\ events 
due to the FB detection asymmetries coming 
from the boost of the center-of-mass system (CMS)
relative to the laboratory.

The production asymmetry has two physical components. 
Interference in $\ep\en\to c\bar{c}$
as mediated by either a virtual $\gamma$ or a virtual $Z^{0}$
contributes at the percent level at this energy, 
and is well understood.
In addition, asymmetries induced by higher-order QED effects are
expected to have polar angle dependence and to
peak sharply in the forward and backward directions~\cite{qedcontrib}.
Although well-considered for $\mu$-pair production~\cite{fbasym},
the precise shape of this contribution for $D$ production
is not known.

We use a data sample corresponding to an integrated luminosity
of 385.8\invfb
collected with the \babar\ detector~\cite{Aubert:2001tu}
at the \pep2\ $e^+ e^-$ collider at SLAC.
The production vertices of charged particles are measured with
a silicon-strip detector (SVT), and their momenta are measured
by the SVT and a drift chamber (DCH) in a 1.5\,T magnetic field.
Information from a Cherenkov-radiation detector,
along with energy-deposition measurements from the SVT and DCH,
provide $K$-$\pi$ discrimination.

We analyze neutral $D$ mesons produced from $\Dstarp\to\Dz\spi^+$;
the charge of the \spi, a low momentum (soft) pion,
indicates the flavor of the \Dz\ at production.
To correct for asymmetry in this flavor tag,
we measure the relative detection efficiency for
soft pions in recorded data using
the decay $\dzkpi$ with (tagged) and without (non-tagged)
soft-pion flavor tagging.  The only detector asymmetry
present in reconstruction of the signal modes is due
to the tagging \spi, since the \CP\ final 
states are reconstructed identically for \Dz\ and \Dzb.

We reconstruct the four decay chains
$\Dz \to K^- \pi^+$; $\Dstarp \to \Dz \spi^+$, $\Dz \to K^- \pi^+$;
$\Dstarp \to \Dz \spi^+$, $\Dz \to K^- K^+$;
and $\Dstarp \to \Dz \spi^+$, $\Dz \to \pi^- \pi^+$.
We require \Dz\ candidates to have
center-of-mass momenta greater than 2.4\gevc, which removes almost
all $B$ decays.  Each \Dz\ daughter must satisfy a likelihood-based
particle-identification selection and must have at least two
position measurements in each of the $z$ and $\phi$ coordinates of
the SVT.  We require $\pi_s^{\pm}$ candidates
to have a lab momentum greater than 100\mevc\
and at least six position measurements in the SVT.

For $h = K,\pi$, we accept candidates with an invariant mass
$1.79 < m_{hh} < 1.93\gevcc$ and, for
final states with a \spi, an invariant mass difference
$0.140 < \dm < 0.152\gevcc$, where
$\dm \equiv m_{hh\pi_{s}} - m_{hh}$.
For each \Dz\ candidate, we constrain the $h^+h^-$ tracks to originate from
a common vertex; for applicable final states,
we also require the \Dz\ and \spi\ to originate from a common vertex
within the $e^+ e^-$ interaction region.
We select candidates for which the $\chi^2$ probability of the vertex fit
of the two \Dz\ daughters is greater than 0.005.
For the $K\!K$ and $\pi\pi$ modes, final asymmetries are calculated
using events for which the polar angle of the
\Dz\ momentum in the CMS with respect
to the beam axis satisfies
$|\cos\theta_{\Dz}^{\textrm{CMS}}| < 0.8$.

\begin{figure}[!t]
\begin{center}
\includegraphics[width=\linewidth]{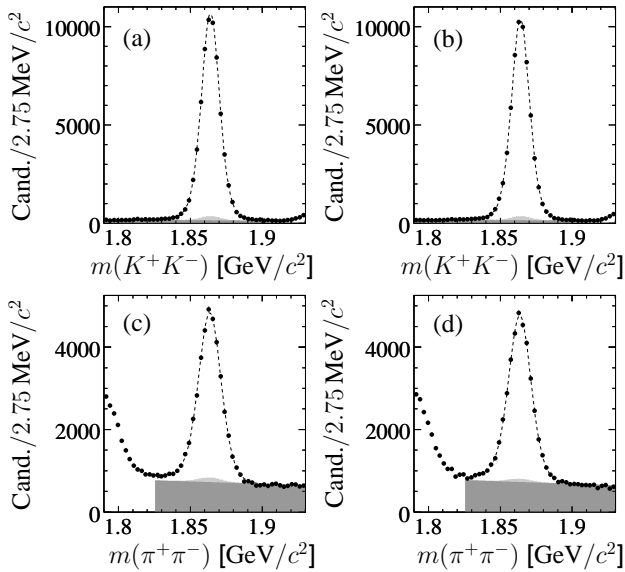}
\caption{
Invariant mass distributions of the $K\!K$ final state
tagged as (a) \Dz\ and (b) \Dzb,
and the $\pi\pi$ final state tagged as (c) \Dz\ and (d) \Dzb.
Distributions of data (points with error bars)
in the signal region $0.1434 < \dm < 0.1474\gevcc$
are overlaid with fitted PDFs (dashed line, shaded areas).
The white regions under the central peaks represent
signal events, the light gray misassociated $\pi_s^{\pm}$ events,
and the dark gray remaining nonpeaking background.
The data are shown over ranges extended beyond the fitted regions
to illustrate the physical background shapes.
\vspace{-5ex}}
\label{fig:datafit}
\end{center}
\end{figure}

We statistically separate signal from background in the selected events
by calculating signal weights based on an optimized likelihood function~\cite{splot}.
The likelihood function is composed of probability density functions (PDFs)
that are fitted to the mass distributions using the maximum likelihood
technique.  For the non-tagged sample, a one-dimensional PDF is fitted to the
$m_{K\pi}$ distribution; for the tagged samples, two-dimensional PDFs are
fitted to the $m_{hh}$ and \dm\ distributions.
Two-dimensional PDFs are used for the tagged samples to account for
possible asymmetries in the background from correctly reconstructed
\Dz\ decays with a misassociated \spi\ candidate; this background category
peaks in $m_{hh}$ but does not peak in \dm.  The PDFs in this analysis
are nearly identical to those used in an analysis of the decay
\dzkpiws~\cite{Aubert:2007wf}, since the signal shapes and background sources are
very similar.  Although the PDFs are motivated by studies of simulated events,
all of the shape parameters are varied in the fits to recorded data.
Our selection of PDFs is treated as a source of systematic uncertainty.
Because the signal shape is indistinguishable for \Dz\ and \Dzb\ distributions,
we use the same signal PDF to describe both flavors of a mode and
fit it to them simultaneously to reduce statistical uncertainties.
The $K\!K$ and $\pi\pi$ invariant mass distributions for \Dz\ and \Dzb,
with fitted PDFs overlaid, are shown in Fig.~\ref{fig:datafit}.
This analysis is sensitive only to ratios of \Dz-signal yields
to \Dzb-signal yields, and not to absolute yields, so the final results
are relatively insensitive to the exact forms of the PDFs.

\begin{table*}[!t]
\caption{Signal yields in reconstructed modes.  Listed uncertainties are statistical only.
Corrections are applied only to \Dz\ samples, but all post-correction samples are restricted
to the phase space of the correction map.}
\begin{center}
\begin{tabular}{crcrrcrcrr}
\hline
\hline
  &
 \multicolumn{6}{c}{Raw yields} &
 \multicolumn{3}{c}{Post-correction yields} \\
 Final state &
 \multicolumn{3}{c}{\Dz} & \multicolumn{3}{c}{\Dzb} &
 Corr. used &
 \multicolumn{1}{c}{\Dz} & \multicolumn{1}{c}{\Dzb} \\
\hline
 $K\pi$ & 3,363,000 & $\pm$ & 6,000 & 3,368,000 & $\pm$ & 6,000 &
   none & \multicolumn{1}{c}{---} & \multicolumn{1}{c}{---} \\
 $K\pi\spi$ & 705,100 & $\pm$ & 1,000 & \hspace{0.5em} 703,500 & $\pm$ & 1,000 &
   $K\pi$ map & $633,300$ & \hspace{0.5em}$630,100$ \\
 $K\!K\spi$ & 65,730 & $\pm$ & 340 & 63,740 & $\pm$ & 330 &
   \spi\ map & $65,210$ & $63,490$\\
 $\pi\pi\spi$ & 32,210 & $\pm$ & 310 & 31,930 & $\pm$ & 310 &
   \spi\ map & $31,900$ & $31,760$\\
\hline
\hline
\end{tabular}
\label{tbl:yields}
\vspace{-3ex}
\end{center}
\end{table*}

\begin{figure}[!t]
\begin{center}
\includegraphics[width=\linewidth]{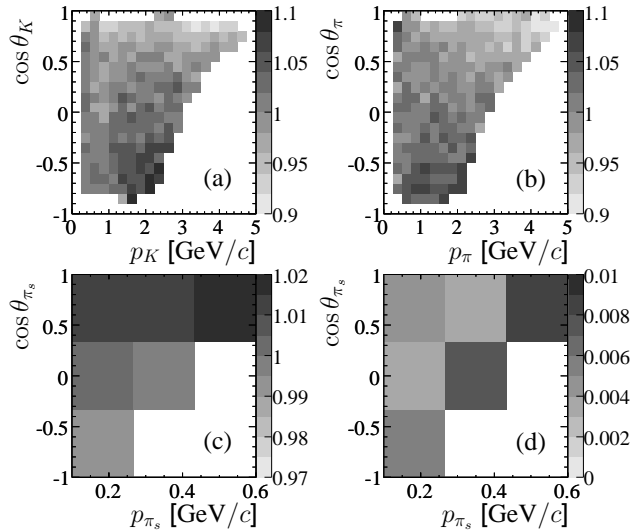}
\caption{
$K\pi$ efficiency-map components obtained from the
non-tagged \Dz\ daughters (a) $K$ and (b) $\pi$,
and (c) \spi\ efficiency map with (d) statistical errors
from the tagged $K\pi$ sample.
Maps are produced from the ratios of
candidate numbers of \Dz\ to \Dzb.
\vspace{-5ex}}
\label{fig:releff}
\end{center}
\end{figure}

The decay \dzkpi\ is chosen as a calibration mode 
because it provides an easily reconstructed
independent sample with high statistics. However,
detector asymmetries in reconstruction of the \Dz\ final state
cannot be ignored (see Fig.~\ref{fig:releff}(a,b)).
These must be corrected to isolate the soft-pion asymmetry.

Using the non-tagged $K\pi$ sample, we produce a map of the relative 
reconstruction efficiency between \Dz\ and \Dzb\ in this final state
in terms of the momenta of both \Dz\ daughters, shown by components in
Fig.~\ref{fig:releff}(a,b).
For each \Dz\ daughter, we consider
the momentum magnitude and polar angle
in the lab with respect to the beam axis; these components are 
correlated. The daughters are, however,
factorizable from one another.  By considering the normalized product of the
$K$ and $\pi$ efficiency-map components, we obtain a four-dimensional
relative-efficiency map for correcting \dzkpi\ relative to \dzbkpi.
The presence of prompt \Dz\ decays not originating
from a \Dstarp\ in the non-tagged sample extends the kinematic boundaries of the
map but does not otherwise affect it.

This $K\pi$ map is used to weight the \Dz\ candidates in
the slow-pion tagged $K\pi$ sample,
eliminating asymmetries due to the \Dz/\Dzb\ daughters. Because all
charm production is subject to the same production 
asymmetries, these are simultaneously removed from the tagged $K\pi$ sample
by this correction. After the weights have been applied,
the remaining asymmetry in the sample is due to the relative
soft-pion efficiency.

We produce a map of the relative soft-pion efficiency
in terms of the pion-momentum magnitude and polar angle
in the lab (Fig.~\ref{fig:releff}(c)).
Charm production is azimuthally uniform, and
$\phi$ is found to be uncorrelated with other momentum variables.
Therefore, the $\phi$ dependence is accounted for by an integrated scale factor.
The uncertainties shown (Fig.~\ref{fig:releff}(d))
are due to the statistical uncertainties in the sample yields.
Signal-mode \Dz\ yields are weighted with this \spi\ map to correct
for the soft-pion tagging asymmetry. 
The signal modes (with remaining production asymmetries) can thus be analyzed for
evidence of \CP\ violation.
In Table~\ref{tbl:yields}, we
list the raw and post-correction yields for the calibration and
signal samples in this analysis.
In calculating these corrections, histogram bins near kinematic boundaries
with fewer than 5,000 events are removed.

\begin{table}[!b]
\caption{Summary of systematic uncertainties.}
\begin{center}
\begin{tabular}{crr}
\hline
\hline
 \rule{0em}{3ex} Category & \multicolumn{1}{c}{$\Delta a_{\CP}^{K\!K}$} & \multicolumn{1}{c}{$\Delta a_{\CP}^{\pi\pi}$} \\
\hline
 \rule{0em}{3ex} 2-Dim. PDF shapes & $\pm 0.04\%$ & $\pm 0.05\%$ \\
 \spi\ correction & $\pm 0.08\%$ & $\pm 0.08\%$ \\
 $a_{\CP}$ extraction & $\pm 0.09\%$ & $\pm 0.20\%$ \\
 \rule{0em}{3ex} Quadrature sum & $\pm 0.13\%$ & $\pm 0.22\%$\\
\hline
\hline
\end{tabular}
\label{tbl:syst}
\vspace{-3ex}
\end{center}
\end{table}

\CP\ violation would appear as an asymmetry in \Dz/\Dzb\
yields, independent of any kinematic variables.
Because of the FB asymmetry in production, we calculate
yield asymmetries as a function of
$\cos\theta = \cos\theta_{\Dz}^{\textrm{CMS}}$ and decompose
these into even and odd parts.
We define
\begin{eqnarray}
 & \displaystyle%
 a^{\pm}(\cos\theta) = \frac{n_{\Dz}(\pm|\cos\theta|) - n_{\Dzb}(\pm|\cos\theta|)}%
 {n_{\Dz}(\pm|\cos\theta|) + n_{\Dzb}(\pm|\cos\theta|)} & \\
 \label{eq:acp}
 & \rule{0em}{3ex}a_{\CP} = a_{\CP}(\cos\theta) \approx %
 \left(a^{+}(\cos\theta) + a^{-}(\cos\theta)\right)/2 & \\
 \label{eq:afb}
 & \rule{0em}{3ex}a_{\textrm{FB}}(\cos\theta) \approx %
 \left(a^{+}(\cos\theta) - a^{-}(\cos\theta)\right)/2\textrm{,} &
\end{eqnarray}
where $n_{\Dz}$ and $n_{\Dzb}$ are the numbers of signal events for
\Dz\ and \Dzb\ after applying the weights discussed above,
$a_{\CP}$ is the even component and $a_{\textrm{FB}}(\cos\theta)$ the odd component.
Eqs.~\ref{eq:acp} and~\ref{eq:afb} are approximate as second-order terms
in $a^{\pm}$ have been omitted.
The even part, representing \CP-violating effects, would provide
evidence of a uniform yield asymmetry. The odd part
represents the production asymmetry, including higher-order QED contributions.
From the several values of $a_{\CP}$ obtained as a function of
$|\cos\theta|$, we obtain a central value from a $\chi^2$ minimization.

\begin{figure}[!t]
\begin{center}
\includegraphics[width=\linewidth]{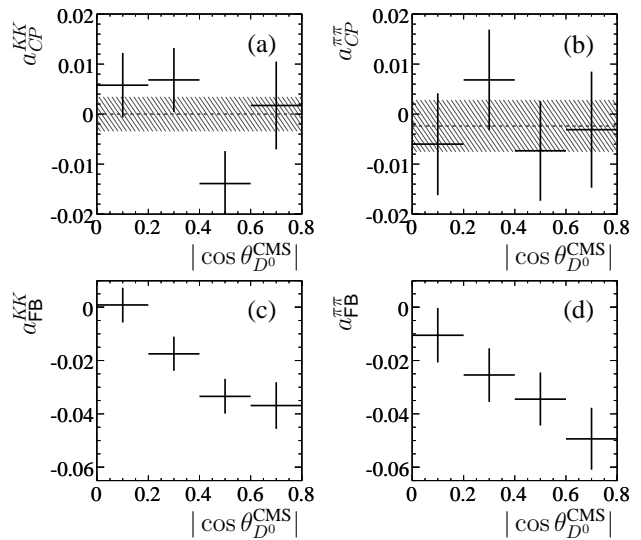}
\caption{
\CP-violating asymmetries in (a) $K\!K$ and (b) $\pi\pi$,
and forward-backward asymmetries in
(c) $K\!K$ and (d) $\pi\pi$.  In (a) and (b), the
dashed lines represent the central values and the
hatched regions the 1$\sigma$ intervals, obtained from
$\chi^2$ minimizations.
\vspace{-5ex}}
\label{fig:asym}
\end{center}
\end{figure}

We consider three sources of systematic error to be significant.
One source is the choice of PDFs used to describe the
signal and background distributions, which affects the statistical
background subtraction.  We estimate this systematic uncertainty
by substituting different background shapes in $m$ and \dm\
and an alternative two-dimensional signal shape in the fits to the
tagged samples.  Another source is the
binning choices made and dependences in the \spi-efficiency correction.
We estimate the size of this uncertainty by varying the number
of bins and the required number of events per bin
in histograms used to calculate efficiencies, and by adding
a $\phi$ dependence to the efficiency correction.  We find the largest
uncertainty here arises from the particular choice of binning
in the \spi-efficiency map.  Because the systematic uncertainty
in applying the \spi-efficiency correction is the same for both modes,
we evaluate its size using the larger signal sample.
Finally, we consider
the procedure for extracting $a_{\CP}$.  We vary the binning
and the accepted range of $|\cos\theta|$; the largest uncertainty
comes from the latter.  All other sources of systematic uncertainty
are highly suppressed because the final states are reconstructed identically
for \Dz\ and \Dzb.  We summarize the contributions to the
total systematic uncertainty in Table~\ref{tbl:syst}.
The smaller $\pi\pi$ sample size influences the calculation of
its systematic uncertainty.

For $K\!K$, we measure
$a_{\CP}^{K\!K} = (0.00 \pm 0.34\,\textrm{(stat.)} \pm 0.13\,\textrm{(syst.)})\%$.
For $\pi\pi$, we measure
$a_{\CP}^{\pi\pi} = (-0.24 \pm 0.52\,\textrm{(stat.)} \pm 0.22\,\textrm{(syst.)})\%$.
Statistical uncertainties of 0.1\% in the \spi\ correction have been included in the final
statistical uncertainty values.
The even and odd asymmetries for each mode as a function of $|\cos\theta|$ are
shown in Fig.~\ref{fig:asym}.
We conclude from the $\chi^2$ minimizations
in Fig.~\ref{fig:asym}(a,b) that there is no
evidence of \CP\ violation
in either of the Cabibbo-suppressed two-body modes of \Dz\ decay.
This result is in agreement with Standard Model predictions.
It also provides a new constraint on theories beyond the Standard
Model~\cite{newmodels}, some of which predict significant levels of 
\CP\ violation in these modes.
The asymmetries observed in Fig.~\ref{fig:asym}(c,d)
represent the two Standard Model asymmetries discussed. Although an exact prediction
of these forward-backward asymmetries 
does not exist, the observed values are consistent with expectations.

We are grateful for the excellent luminosity and machine conditions
provided by our \pep2\ colleagues, 
and for the substantial dedicated effort from
the computing organizations that support \babar.
The collaborating institutions wish to thank 
SLAC for its support and kind hospitality. 
This work is supported by
DOE
and NSF (USA),
NSERC (Canada),
CEA and
CNRS-IN2P3
(France),
BMBF and DFG
(Germany),
INFN (Italy),
FOM (The Netherlands),
NFR (Norway),
MIST (Russia),
MEC (Spain), and
STFC (United Kingdom). 
Individuals have received support from the
Marie Curie EIF (European Union) and
the A.~P.~Sloan Foundation.

\end{document}